\documentclass[a4paper]{paper}
\usepackage{amsmath}
\usepackage{amssymb}
\usepackage{amsthm} 
\usepackage{mathrsfs}
\usepackage{amsbsy} 
\usepackage{mathtools}

\usepackage{memhfixc}
\usepackage{latexsym}
\usepackage{chemarrow} 
\usepackage{bm}
\usepackage{bbm}
\usepackage{turnstile}

\usepackage{nicefrac}
 
\usepackage{braket} 

\usepackage{graphicx}
\graphicspath{{grafici/},{foto/}} 
\usepackage{wrapfig}
\usepackage{float}
\usepackage{lpic} 
\usepackage[small,bf]{caption}
\usepackage{subcaption}
\usepackage{longtable}
\usepackage{rotating}
\usepackage[a4paper]{geometry}
\usepackage[cm]{fullpage}
\usepackage{dcolumn,booktabs}

\usepackage{lmodern} 
\usepackage{type1ec}
\usepackage[T1]{fontenc}
\usepackage{lettrine} 
\usepackage{verbatim}
\usepackage{titling}
\usepackage{microtype}
\usepackage{sidecap}
\usepackage{frontespizio}
\usepackage{xcolor} 
\usepackage{empheq} 
\usepackage{layaureo} 		
\usepackage{fancyhdr}		
\usepackage[italian]{varioref}    

\usepackage[latin1]{inputenc}
\usepackage[english]{babel}

\newcommand{\deriv}[2]{\dfrac{\mathrm{d} #1}{\mathrm{d} #2}}
\newcommand{\de}[0]{\textrm{d}}
\newcommand{\dis}[1]{\displaystyle{#1}}
\newcommand{\en}{\mathrm{e}}
\newcommand{\ii}{\mathrm{i}}
\newcommand{\pderiv}[2]{\dfrac{\partial #1}{\partial #2}}

\newcommand{\q}[1]{\left[#1\right]}
\newcommand{\ton}[1]{\left(#1\right)}
\newcommand{\funz}[1]{\hspace{-.06cm}\left(#1\right)}
\newcommand{\g}[1]{\left\{#1\right\}}

\newcommand{\im}[0]{\textrm{i}}

\renewcommand{\vec}[1]{{\bm#1}}

\usepackage[autostyle,italian=guillemets,]{csquotes}
\usepackage[bibstyle=numeric, sorting=none, backend=biber]{biblatex}
\bibliography{ob.bib}

\title{The Born Oscillator}
\author{Gianni Coppa\\ Dipartimento di Elettronica e Telecomunicazioni\\Politecnico di Torino\\Torino, Italy}
\date{}

\begin{document}
\maketitle
\begin{abstract}
The paper studies the properties of an oscillator whose Hamiltonian is ${\q{\ton{1+q^2}\ton{1+p^2}}^{1/2}-1}$. It can be deduced from the nonlinear theory of electrodynamics originally proposed by Max Born in 1934. The quantization of such oscillator represents a possible regularization of the Barry and Keating's Hamiltonian, which has been proposed in the framework of the theory of non-trivial zeros of the Riemann's $\zeta$ function.
\end{abstract}
\section{Introduction}
The theory of electromagnetism proposed by Max Born in 1934 \cite{Born_solo} had as its main purpose a modification of the classic Maxwell equations in order to avoid field singularities in the presence of point sources. The Lagrangian density of the electromagnetic field, which according to Maxwell's theory is
%
$\mathscr{L}_M=-\dfrac{1}{4}F_{\mu\nu}F^{\mu\nu},$
%
where $F^{\mu\nu}=\partial^\mu A^\nu-\partial^\nu A^\mu$ is the Faraday tensor, is here replaced by:
\begin{equation}\label{eq:intro_lag_den_born}
\mathscr{L}_B=\dfrac{1}{\varepsilon^2}\g{1-\ton{1-2\varepsilon^2\mathscr{L}_M}^{\frac{1}{2}}},
\end{equation}
where $\varepsilon$ is a parameter that allows to fix the maximum value of the electric field (for $\varepsilon \rightarrow0$, $\mathscr{L}_B$ is reduced to $\mathscr{L}_M$).
Born's theory, which from the point of view of physics is mainly of historical interest (having been replaced by modern quantum electrodynamics), has however been the object of numerous studies \cite{Barbashov:1966frq, barbashov1966, BialynickiBirula:1984tx, Chernitskii2005}, as it presents mathematical peculiarities of considerable interest. The object of the present work is the analysis of a particular type of oscillator, here referred to as the Born oscillator, which in the aforementioned theory replaces the harmonic oscillator of classical field theories as a fundamental element. The properties of this oscillator are discussed in detail in Sect. \ref{sec:prop}. The Sect. \ref{sec:prob} deals with the problem of the quantization of the Born oscillator, which presents analogies with the theory of Berry and Keating \cite{BK99a, BK99b, Berry_2011, Srednicki_2011} relating to the Hilbert's hypothesis concerning the Riemann's $\zeta$ function, according to which the imaginary part of the non-trivial $\zeta$'s zeros are the eigenvalues of a suitable Hermitian operator.
\section{The Born oscillator}\label{sec:prop}
Here we consider a very simple case, where $\mathscr{A}^\mu=(0,0,0,\mathscr{A}_z)$, for which the Born Lagrangian density becomes:
\begin{equation}\label{eq:prop_1}
\mathscr{L}_B\funz{\mathscr{A}_z,\pderiv{\mathscr{A}_z}{t}}=\dfrac{1}{\varepsilon^2}\g{1-\q{1-\varepsilon^2\ton{\pderiv{\mathscr{A}_z}{t}}^2+\varepsilon^2\ton{\pderiv{\mathscr{A}_z}{x}}^2}^{\frac{1}{2}}}.
\end{equation}
The momentum density associated with it is:
\begin{equation}\label{eq:prop_2}
\Pi_z=\pderiv{\mathscr{L}_B}{\ton{\partial \mathscr{A}_z/\partial t}}=\dfrac{\partial \mathscr{A}_z/\partial t}{\q{1-\varepsilon^2\ton{\pderiv{\mathscr{A}_z}{t}}^2+\varepsilon^2\ton{\pderiv{\mathscr{A}_z}{x}}^2}^{\frac{1}{2}}},
\end{equation}
or:
\begin{equation}\label{eq:prop_3}
\pderiv{\mathscr{A}_z}{t}=\Pi_z\cdot\q{\dfrac{1+\varepsilon^2\ton{\pderiv{\mathscr{A}_z}{x}}^2}{1+\varepsilon^2\Pi_z^2}}^{\frac{1}{2}}.
\end{equation}
The Hamiltonian density is therefore:
\begin{equation}\label{eq:prop_4}
\begin{array}{rcl}
\mathscr{H}&=&\Pi_z\pderiv{\mathscr{A}_z}{t}-\mathscr{L}_B=\vspace{.3cm}\\
&=&\dfrac{1}{\varepsilon^2}\g{\q{\ton{1+\varepsilon^2\Pi_z^2}\ton{1+\varepsilon^2\ton{\pderiv{\mathscr{A}_z}{x}}^2}}^{\frac{1}{2}}-1}.
\end{array}
\end{equation}
For $\varepsilon\rightarrow0$, $\mathscr{H}$ tends to $\dfrac{\Pi_z^2+\ton{\partial \mathscr{A}_z/\partial x}^2}{2}$ (i.e., $\dfrac{\vec{E}^2+\vec{B}^2}{2}$, which represents the energy density of the electromagnetic field according to the Maxwell's theory). If the total Hamiltonian $H=\int\mathscr{H}\de {x}$ is spatially discretized as $\dis{\sum_\alpha}\mathscr{H}\funz{{x}_\alpha}\Delta {x}$, in the classical case we will have $\mathscr{H}\funz{{x}_\alpha}\simeq\dfrac{\Pi_{z,\alpha}^2+\ton{A_{z,\alpha}-A_{z,\alpha-1}}^2}{2}$ and, therefore, the physical system described by $\mathscr{H}$ can be regarded as a sequence of coupled  harmonic oscillators. Since the Hamiltonian of a single harmonic oscillator (in suitable units), is simply $\dfrac{p^2+q^2}{2}$, in the case of Born theory it must be replaced by:
\begin{equation}\label{eq:prop_5}
\mathscr{H}_B\funz{q,p}=\dfrac{1}{\varepsilon^2}\g{\q{\ton{1+\varepsilon^2 p^2}\ton{1+\varepsilon^2 q^2}}^{\frac{1}{2}}-1}.
\end{equation}
The new Hamiltonian describes a particular type of oscillator, which is referred as the Born oscillator in the following. Its equations of motion, written in terms of $\tilde{q}=\varepsilon q$, $\tilde{p}=\varepsilon p$, are:
%
\begin{equation}\label{eq:prop_6}
\deriv{\tilde{q}}{t}=\tilde{p}\ton{\dfrac{1+\tilde{q}^2}{1+\tilde{p}^2}}^{\frac{1}{2}},\quad\deriv{\tilde{p}}{t}=-\tilde{q}\ton{\dfrac{1+\tilde{p}^2}{1+\tilde{q}^2}}^{\frac{1}{2}}.
\end{equation}
Obviously, for small values of $\tilde{q}$ and $\tilde{p}$ (i.e., $|q|,|p|\ll1/\varepsilon$) one has $\mathscr{H}_B\simeq\dfrac{p^2+q^2}{2}$ and Eqs.\eqref{eq:prop_6} reduce to the equations of motion for a harmonic oscillator. In general, the system described by Eqs. \eqref{eq:prop_6} evolves in time describing closed trajectories in the phase space such that $\mathscr{H}_B=\mathscr{E}$, i.e.:
\begin{equation}\label{eq:prop_7}
\ton{1+\tilde{q}^2}\ton{1+\tilde{p}^2}=\ton{1+\varepsilon^2\tilde{\mathscr{E}}}^2.
\end{equation}
For $\varepsilon^2\mathscr{E}\ll1$, the trajectories are close to circles ($\tilde{q}^2+\tilde{p}^2=2\varepsilon^2\mathscr{E}$), while for $|q|,|p|\gg 1/\varepsilon$  they tend to branches of hyperbola:
\begin{equation}\label{eq:prop_8}
|\tilde{p}|\cdot|\tilde{q}|\simeq\varepsilon^2\mathscr{E}.
\end{equation}
%
%
It is  simple to evaluate the period $T$ of oscillation as a function of the energy $\mathscr{E}$. In fact, by eliminating $\tilde{p}$ between the first and second Eq. \eqref{eq:prop_6}, we will have:
\begin{equation}\label{eq:prop_10}
\ton{\deriv{\tilde{q}}{t}}^2=\dfrac{1+\tilde{q}^2}{1+\tilde{q}_M^2}\ton{\tilde{q}_M^2-\tilde{q}^2},
\end{equation}
where $\tilde{q}_M$ is the maximum value of $\tilde{q}$ in the oscillation, i.e., $1+\tilde{q}_M^2=\ton{1+\varepsilon^2\tilde{\mathscr{E}}}^2$. Posing $\tilde{q}=\tilde{q}_M\cos\funz{\theta}$, from Eq. \eqref{eq:prop_10} we immediately obtain:
\begin{equation}\label{eq:prop_11}
\dfrac{T}{4}=\dis{\int_0^{\pi/2}\ton{1-\dfrac{\tilde{q}^2_M}{1+\tilde{q}^2_M}\sin^2\funz{\theta}}^{-\frac{1}{2}}\de\theta}=\mathbb{K}\funz{1-\dfrac{1}{\ton{1+\tilde{\mathscr{E}}}^2}},
\end{equation}
where $\mathbb{K}$ is the complete elliptic function of the second kind. For $\varepsilon\rightarrow0$ the period tends to $4\mathbb{K}\funz{0}=2\pi$, as was to be expected (limit for small oscillations). Let us now consider the opposite situation, in which $\varepsilon,\tilde{q}_M\gg$ 1. Assuming $\tilde{q}\funz{t=0}=\tilde{q}_M$, in the time interval $\q{-{T}/{8},{T}/{8}}$ it will always result in $\tilde{q}\gg1$, and therefore the equations of motion can be approximated as:
\begin{equation}\label{eq:prop_12}
\deriv{\tilde{q}}{t}=\dfrac{\tilde{q}\tilde{p}}{\ton{1+\tilde{p}^2}^{{1}/{2}}},\quad\deriv{\tilde{p}}{t}=-\ton{1+\tilde{p}^2}^{\frac{1}{2}}.
\end{equation}
Since the second equation \eqref{eq:prop_12} does not contain $\tilde{q}$, it is straightforward to find the solution of the system, as:
\begin{equation}\label{eq:prop_13}
\tilde{q}\funz{t}=\dfrac{\tilde{q}_M}{\cosh\funz{t}},\quad \tilde{p}\funz{t}=-\sinh\funz{t}.
\end{equation}
Given the symmetry of Eqs. \eqref{eq:prop_6} with respect to $\tilde{p}$ and $\tilde{q}$, in the interval $\q{-{3T}/{8},{T}/{8}}$, where $\tilde{p}\funz{t}\gg1$, it will result:
\begin{equation}\label{eq:prop_14}
\tilde{q}=\sinh\funz{t+{T}/{4}},\quad \tilde{p}=\dfrac{\tilde{q}_M}{\cosh\funz{t+{T}/{4}}}.
\end{equation}
For $t=-T/8$ the two solutions will both be valid, being $\tilde{q}\funz{-T/8}\gg1$ and $\tilde{p}\funz{-T/8}\gg1$, and therefore it must hold:
\begin{equation}\label{eq:prop_15}
\dfrac{\tilde{q}_M}{\cosh\funz{T/8}}=\sinh\funz{T/8},
\end{equation}
or
\begin{equation}\label{eq:prop_16}
{T}={4}\sinh^{-1}\funz{2\tilde{q}_M}.
\end{equation}
The previous calculation and the symmetry of Eqs. \eqref{eq:prop_6} by exchanging  $\tilde{p}$ and $\tilde{q}$ suggest the possibility of using two new dynamic variables, $\mathbb{P}$ and $\mathbb{Q}$, such that: 
\begin{equation}
\tilde{p}=\sinh\funz{\mathbb{P}},\quad \tilde{q}=\sinh\funz{\mathbb{Q}},
\end{equation}
 for which the equations of motion are:
\begin{equation}\label{eq:prop_17}
\deriv{\mathbb{Q}}{t}=\tanh\funz{\mathbb{P}},\quad \deriv{\mathbb{P}}{t}=-\tanh\funz{\mathbb{Q}}.
\end{equation}
They can be considered as deriving from the new Hamiltonian:
\begin{equation}\label{eq:prop_17b}
\mathbb{H}_B\funz{\mathbb{Q},\mathbb{P}}=\log\q{\cosh\funz{\mathbb{P}}\cosh\funz{\mathbb{Q}}}.
\end{equation}
It is interesting to consider the case in which a forcing term is added to the Born Hamiltonian:
\begin{equation}\label{eq:prop_18}
H\funz{q,p}=\mathscr{H}_B\funz{q,p}-F\cdot q
\end{equation}
(it corresponds to the term ${j}_z\mathscr{A}_z$ in the Hamiltonian density). For $|\tilde{q}|,|\tilde{p}|\ll1$ it reduces to the classical Hamiltonian of a harmonic oscillator subjected to the force $F$. In that case, there is a stationary solution, $q=F$,  $p=0$ and in general the motion consists of oscillations with respect to this equilibrium point. In the case of the Hamiltonian \eqref{eq:prop_18}, the equations of motion being:
\begin{equation}\label{eq:prop_19}
\deriv{\tilde{q}}{t}=\tilde{p}\ton{\dfrac{1+\tilde{q}^2}{1+\tilde{p}^2}}^{\frac{1}{2}}, \deriv{\tilde{p}}{t}=-\tilde{q}\ton{\dfrac{1+\tilde{p}^2}{1+\tilde{q}^2}}^{\frac{1}{2}}+\varepsilon F
\end{equation}
 it is immediate to verify the stationary solution:
\begin{equation}
\dfrac{\tilde{q}}{\ton{1+\tilde{q}^2}^{\frac{1}{2}}}=\varepsilon F,\quad\tilde{p}=0.
\end{equation}
However, unlike the classic case, this solution exists only if $|F|<1/\varepsilon$. This is essentially the reason why there is not a stationary solution of the Born's equations when the current density exceeds a critical value \cite{tesi_coppa}.
\section{Quantization of the Born oscillator}\label{sec:prob} 
The problem of the quantization of the Born oscillator appears interesting for several aspects. The first is the peculiar form of the Hamiltonian function, in which the square root of a polynomial in $p$ and $q$ appears. The second reason of interest is that Berry and Keating's theory concerning the non-trivial zeros of Riemann's $\zeta$ function is based upon the regularization of the Hamiltonian $p\cdot q$, to which Eq.\eqref{eq:prop_5} reduces  for $p,q\gg 1/\varepsilon$. As regards the first aspect, instead of transforming $\mathscr{H}_B$ directly into an operator, one can consider a function of it, namely:
\begin{equation}
B=\mathscr{H}_B\cdot\ton{1+\frac{\varepsilon^2}{2}\mathscr{H}_B}.
\end{equation}
In fact, taking the square of Eq.\eqref{eq:prop_5} we have:
\begin{equation}
\ton{1+\varepsilon^2\mathscr{H}_B}^2=1+\varepsilon^2\ton{q^2+p^2}+\varepsilon^4q^2p^2,
\end{equation}
and, consequently,
\begin{equation}
B\funz{p,q}=\frac{1}{2}\ton{q^2+p^2}+\frac{\varepsilon^2}{2}p^2q^2.
\end{equation}
The function $B\funz{p,q}$ can be transformed into a quantum operator $\hat{\mathbb{B}}$, as the sum of the Hamiltonian of the harmonic oscillator, $\frac{1}{2}\ton{\hat{p}^2+\hat{q}^2}$, and of a suitable symmetrized version of $\frac{1}{2}\varepsilon^2\hat{p}^2\hat{q}^2$, e.g., $\frac{1}{2}\varepsilon^2\hat{q}\hat{p}^2\hat{q}$.
In terms of creation and annihilation operators, $\im \hat{p}\hat{q}$ can be written as $\frac{1}{2}\ton{\hat{a}^{\dagger2}-\hat{a}^2+1}$, and consequently
\begin{equation}
\hat{q}\hat{p}^2\hat{q}=\ton{\im\hat{p}\hat{q}}^\dagger\ton{\im\hat{p}\hat{q}}=\frac{1}{4}\q{1-\ton{\hat{a}^{\dagger4}+\hat{a}^4+\hat{a}^{\dagger2}\hat{a}^2+\hat{a}^{\dagger2}\hat{a}^2}}.
\end{equation}
Therefore, if $\ket{n}$ is the$n-$th eigenstate of the harmonic oscillator, one has:
\begin{equation}
\hat{\mathbb{B}}\ket{n}=\ton{n+\frac{1}{2}+\varepsilon^2u_n}\ket{n}-\varepsilon^2v_n\ket{n+4}-\varepsilon^2 v_{n-4}\ket{n-4}.\quad n=0,1,2,...
\end{equation}
being
\begin{equation}
u_n=\frac{1}{8}+\frac{1}{4}\ton{1+n+n^2},\quad v_n=\q{\ton{n+1}\ton{n+2}\ton{n+3}\ton{n+4}}^{1/2}.
\end{equation}
By writing a generic eigenstate of $\hat{\mathbb{B}}$ as
\begin{equation}
\ket{\mathbb{B}'}=\dis{\sum_{n=0}^{+\infty} c_n \ket{n}}
\end{equation}
one has the following recursive relationship for the $c_n$ coefficients:
\begin{equation}\label{eq:cond_K_cn}
\varepsilon^2v_n c_{n+4}-\ton{\frac{1}{2}+\varepsilon^2u_n-\mathbb{B}'}c_n+\varepsilon^2v_{n-4}c_{n-4}=0
\end{equation}
and $\mathbb{B}'$ is an eigenvalue of $\mathbb{B}$ if
\begin{equation}
\dis{\lim_{n\rightarrow\infty}}c_n=0.
\end{equation}
It can be noticed that Eq. \eqref{eq:cond_K_cn} connects the coefficients $c_{m+4n}$, with $m=0,1,2,3$. That means that the eigenvalue spectrum of $\hat{\mathbb{B}}$ is formed by four separate sets of values, each starting from a different initial condition for the first values of $c_n$ (i.e., $c_0=1$, $c_1=c_2=c_3=0$, or $c_1=0$, $c_0=c_2=c_3=0$, and so on). \\
A possible alternative is to define a quantum operator using Weyl's correspondence principle \cite{Weyl}, according to which, starting from a "classical" function $A_c\funz{p,q}$, one first takes its Fourier transform:
\begin{equation}\label{eq:ob27}
a\funz{\xi,\eta}=\frac{1}{2\pi}\iint\de p\,\de q\,\, A_c\funz{p,q}\exp\q{-\ii\ton{\xi p+\eta q}},
\end{equation}
and from this a quantum operator $\hat{A}$ is defined, as:
\begin{equation}\label{eq:ob28}
\hat{A}\funz{\hat{p},\hat{q}}=\frac{1}{2\pi}\iint\de \xi\, \de \eta\,\, a\funz{\xi,\eta}\exp\q{\ii\ton{\xi\hat{p}+\eta\hat{q}}}.
\end{equation}
If this method is used, starting from the Hamiltonian $\mathscr{H}_B$  Eqs. (\ref{eq:ob27},\ref{eq:ob28}) define the quantum Hamiltonian. Furthermore, Weyl's theory allows for an asymptotic evaluation of its eigenvalues $\mathscr{E}_n$ for $n\rightarrow\infty$, simply as:
\begin{equation}\label{eq:cond29}
\displaystyle{\iint_{H\funz{p,q}\le \mathscr{E}_n}\de p\, \de q\,= 2\pi\ton{n+\frac{1}{2}}},
\end{equation}
[natural units are used here, i.e., $\q{\hat{q},\hat{p}}=\ii$], or, equivalently, as:
\begin{equation}\label{eq:cond30}
\displaystyle{\iint_{\mathscr{D}_n}}\de\tilde{p}\,\de\tilde{q}\,=2\pi \varepsilon^2\ton{n+\frac{1}{2}},
\end{equation}
with $\mathscr{D}_n=\g{\ton{\tilde{p},\tilde{q}}:\ton{1+\tilde{q}^2}\ton{1+\tilde{p}^2}\le1+\tilde{q}_n^2}$ and $\ton{1+\tilde{q}_n^2}=\ton{1+\varepsilon^2 \mathscr{E}_n}^2$. 
By exploiting the symmetry of the integration domain, the integral appearing in ${\textrm{Eq. } \eqref{eq:cond30}}$ can be written  as $8 I_n$, being:
\begin{equation}\label{eq:int31}
I_n=\dis{\int_0^{\varepsilon{\mathscr{E}^{1/2}_n}}\de\tilde{q}\,\int_{\tilde{q}}^{\tilde{p}\,\funz{q}}\de \tilde{p}},
\end{equation}
and with
\begin{equation}
\ton{1+\tilde{q}^2}\ton{1+\tilde{p}^2\funz{\tilde{q}}}=1+\tilde{q}_n^2.
\end{equation}
For $n\gg1$, when $\tilde{q}_n\gg1$, we will have $\tilde{q}_n\simeq\varepsilon^2 \mathscr{E}_n$. In addition, for $\tilde{q}\in\q{0,\varepsilon\mathscr{E}_n^{1/2}} $ one has  $\tilde{p}\funz{\tilde{q}}\gg1$, and, consequently,
\begin{equation}
\tilde{p}\funz{\tilde{q}}\simeq\dfrac{\tilde{q}_n}{\sqrt{1+\tilde{q}^2}}.
\end{equation}
Thus the integral \eqref{eq:int31} can be evaluated as:
\begin{equation}
I_n \simeq \dis{\int_0^{{\tilde{q}^{1/2}_n}}\ton{\dfrac{\tilde{q}_n}{\sqrt{1+\tilde{q}^2}}-\tilde{q}} \de\tilde{q} }=\tilde{q}_n\q{\sinh^{-1}\funz{\tilde{q}^{1/2}_n}-\frac{1}{2}}
\end{equation}
For $\tilde{q}_n\gg1$, one has $\sinh^{-1}\funz{\tilde{q}^{1/2}_n}\simeq\frac{1}{2}\log\funz{4{\tilde{q}_n}}$, and finally the condition \eqref{eq:cond29} can be written as\footnote{In fact, the integral \eqref{eq:cond30} can be calculated exactly in terms of complete elliptic functions $\mathbb{K}$ and $\mathbb{E}$, so obtaining: $4\ton{1+\tilde{q}_n^2}\mathbb{K}\funz{-\tilde{q}_n^2}-\mathbb{E}\funz{-\tilde{q}_n^2}=2\pi\varepsilon^2$.}:
\begin{equation}\label{eq:35}
4\tilde{q}_n\q{\log\funz{4\tilde{q}_n}-1}=2\pi\varepsilon^2\ton{n+\frac{1}{2}},
\end{equation}
or, in terms of $\mathscr{E}_n$,
\begin{equation}
\mathscr{E}_n\q{\log\funz{4\varepsilon^2 \mathscr{E}_n}-1}=\frac{\pi}{2}\ton{n+\frac{1}{2}}.
\end{equation}
These formulae show clear analogies with the asymptotic expression for the $N$-th non-trivial zero, $1/2+\ii y_N$, of Riemann's $\zeta$ function  \cite{titchmarsh1986theory,edwards2001riemann}:
\begin{equation}
\dfrac{y_N}{2\pi}\g{\log\funz{\dfrac{y_N}{2\pi}}-1}=N-\frac{7}{8}-\mathscr{S},
\end{equation}
with $\mathscr{S}\simeq\frac{1}{2}$. In particular, for $\varepsilon^2=\frac{1}{8}\pi$, the r.h.s.  of Eq.  \eqref{eq:35} is simply  $\frac{1}{4}n+\frac{1}{8}$ and therefore, if $n=4N-6$, $8\pi q_n$ represents an estimate of $y_N$.\\
Instead, if Hamiltonian \eqref{eq:prop_17b} is employed, condition \eqref{eq:cond30} is replaced by:
\begin{equation}
\dis{\iint_{\mathbb{H}\funz{\mathbb{P},\mathbb{Q}}\le\mathscr{E}_n}}\de \mathbb{P}\de \mathbb{Q}= 2\pi \ton{n+\frac{1}{2}}\varepsilon^2
\end{equation}
(here considering $\q{\hat{\mathbb{Q}},\hat{\mathbb{P}}}=\ii\varepsilon^2$). Again, the integral can be written as $8\mathbb{I}_n$, being:
\begin{equation}
\mathbb{I}_n=\dis{\int_0^{\overline{\mathbb{Q}}_n}}\q{\mathbb{P}\funz{\mathbb{Q}}-\mathbb{Q}}\de\mathbb{Q},
\end{equation}
with
\begin{equation}
\mathbb{P}\funz{\mathbb{Q}}=\cosh^{-1}\q{\exp\funz{\mathscr{E}_n}-\cosh\funz{\mathbb{Q}}},
\end{equation}
and with $\overline{\mathbb{Q}}_n=\cosh^{-1}\q{\exp\funz{\mathscr{E}_n/2}}$, such that $\mathbb{H}\funz{\overline{\mathbb{Q}}_n,\overline{\mathbb{Q}}_n}=\mathscr{E}_n$. By defining $\mathbb{Q}_n$ as the maximum value of $\mathbb{Q}$, i.e., $\mathbb{Q}_n=\cosh^{-1}\q{\exp\funz{\mathscr{E}_n}}$, for $n\gg1$, and consequently when $\mathbb{Q},\overline{\mathbb{Q}}, \mathbb{P}\funz{\mathbb{Q}}\gg1$, it will result:
\begin{equation}
\begin{array}{rcl}
\mathscr{E}_n&=&\mathbb{Q}_n-\log 2,\quad \overline{\mathbb{Q}}_n=\frac{1}{2}\ton{\mathbb{Q}_n+\log 2},\vspace{.3cm}\\
\mathbb{P}\funz{\mathbb{Q}}&=&\mathbb{Q}_n+\log 2-\mathbb{Q}-\log\funz{1+\en^{-2\mathbb{Q}}}.
\end{array}
\end{equation}
Therefore, $\mathbb{I}_n$ can be evaluated as:
\begin{equation}
\mathbb{I}_n=\dis{\int_0^{\overline{\mathbb{Q}}_n}}\q{\mathbb{Q}_n+\log 2-2\mathbb{Q}-\log\funz{1+\en^{-2\mathbb{Q}}}}\de\mathbb{Q}=\frac{1}{4}\ton{\mathbb{Q}_n+\log2}^2-\dfrac{\pi^2}{24}
\end{equation}
having considered 
\begin{equation}
\dis{\int_0^{\overline{\mathbb{Q}}_n}}\log\funz{1+\en^{-2\mathbb{Q}}}\simeq\dis{\int_0^{+\infty}\ton{\en^{-2\mathbb{Q}}-\frac{1}{2}\en^{-4\mathbb{Q}}+\frac{1}{3}\en^{-6\mathbb{Q}}-...}}\de\mathbb{Q}=\dfrac{\pi^2}{24}.
\end{equation}
Finally, we have:
\begin{equation}
\ton{\mathscr{E}_n+2\log 2}^2=\dfrac{\pi^2}{6}+\pi\varepsilon^2\ton{n+\frac{1}{2}}.
\end{equation}
\section{Concluding remarks and acknowledgements}
The author came across this peculiar type of oscillator while working on a different subject, concerning the propagation of solitons according to the Born's electrodynamics.  He is grateful to prof. Roberto Tateo for suggesting him the subject of the present research and for the continuous  advice during the work.  The author hopes the results that are presented in the paper will be of interest for the experts of Berry-Keating theory and may provide a contribution on the subject.
\printbibliography
\end{document}